# Development of organic passive devices for RF applications

*A. Aliane, A. Daami, T. Card, R. Gwoziecki, I. Chartier, R. Coppard*
*CEA LITEN, DTNM, 17 rue des martyrs Grenoble, 38054 Cedex 9, France*

*E. Bergeret, E. Bènevent, P. Pannier*
*IM2NP, IMT Technopole Château-Gombert, 13451 Marseille Cedex 20, France*

**Abstract**

*A huge interesting progress in the field of organic electronic materials and devices has been observed in the last decade. However, the understanding of these materials is still a challenge to overcome. Most studies in literature focus on active devices such as OTFTs, OLEDs and OPVs. Nevertheless, a complete technology has to have also passive devices in order to allow the design of interesting applications and complex circuits.*

*This paper deals with the development of a complete set of passive devices allowing the fabrication of simple applications such as filters or sensors. The process flow is a fully screen printed technology that uses exclusively organic materials on gold laser ablated flexible substrate. Discrete passive (R, L, C) devices have been processed and characterized. This has permitted the fabrication of RLC low-band pass filters that are dedicated to RF applications, typically around 1GHz. Furthermore, based on these discrete passive components, we have developed a sensitive sensor on flexible substrate for RFID applications. We present the state of the art of our process development for RF applications using organic materials.*

## 1. Introduction

In recent years, several developments in the field of organic components emerged and, in particular, the realization of thin film transistors (TFT) on flexible substrate was carried out [1] [2]. Nevertheless the fabrication of passive components and their integration in complex circuits still needs the development and the optimization of different organic materials. Moreover the integration of these materials on flexible substrates is sometimes complicated due to the low glass transition temperature (Tg) of the substrate materials such as Polyethylene Naphtalate (PEN) and Polyethylene Terephtalate (PET). Indeed these two substrates have been widely used for the development of organic technologies because of their favourable mechanical properties [3].

To explore our material characteristics in the field of the RF applications, we have fabricated and characterized discrete R, L, C components dedicated to temperature sensing and RF devices integration. In a second step, we have designed and fabricated by the screen-printing technique RLC filters and temperature sensors on PEN substrates. This paper details the process fabrication and the first results of these RLC filters and temperature sensor.

## 2. Technology and integration

### 2.1 RLC filters

The RLC filters are fabricated on a 125 µm thick PEN (polyethylene naphtalate) flexible substrate. The substrate is provided with a 30 nm thick evaporated gold (Au) layer used as the first conducting layer. The bottom Au electrodes of the RLC filters are defined by laser ablation. The electrical capacitor is made of a fluoropolymer dielectric material with a low relative dielectric constant ($\varepsilon_r = 2$). This dielectric material was deposited by the screen-printing technique then annealed at 100°C for 30 minutes. The average thickness of the dielectric in our case is around 2 µm. The resistor and the inductor components are then realized by screen-printing using a silver conductor paste with a good electrical conductivity, on the dielectric layer. Finally, the RLC organic filters are passivated using a thick insulating layer, and keeping appropriate openings for the probe contacting. Chen *et al.* have used the inkjet printing technique for the fabrication of their RC filters [4]. The designed RLC filter has an area of 84 mm$^2$ with adapted PADs for RF measurements.

Figure 1 shows respectively the layout and a top view picture of the realized RLC filters. Different silver pastes have been tested for our RLC filters.

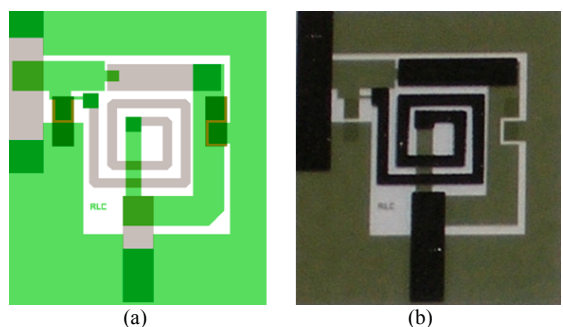

(a)  (b)
*Figure 1.* Layout (a) and image (b) of the RLC filter

Table 1 summarizes the characteristics of our processed RLC filters and their differences.

| Filter number | Bottom electrode | Dielectric | Top electrode | Solvent | Boiling point |
|---|---|---|---|---|---|
| 1 | Au | Fluoropolymer | Ag paste | Toluene | 110°C |
| 2 | Au | Fluoropolymer | Ag Nanoparticles | α-terpineol | 217°C |

*Table 1.* Characteristics of the fabricated RLC filters

## 2.2 Experimental results and modelling

High-frequency measurements of the RLC filters have been carried out using a vector network analyzer (ZVA 24 Rohde & Schwarz) and a classical SOLT (Short-Open-Load-Thru) calibration was applied. A ceramic chuck is also utilized to avoid bringing a ground plane under the flexible substrate and eliminating parasitic capacitors.

From the S21 parameter measurements we have determined the gain and the phase of the two different RLC filters. We show on figure 2 the corresponding Bode-plots of both filters.

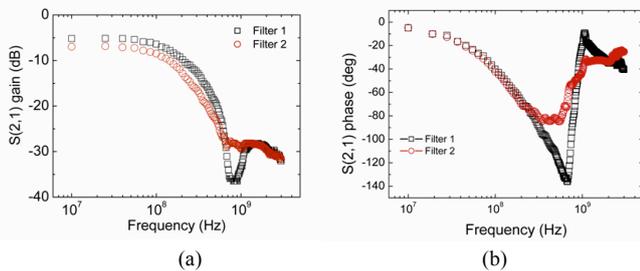

*Figure 2.* Transmission S21 parameters: gain (a) and phase (b) of the filters

At a first glance we can see that both filters show high cut-off frequencies ($f_{-3dB}$). Nevertheless, filter 1 shows a slightly higher $f_{-3dB}$ than filter 2 (155 MHz compared to 137 MHz). This difference is explained from our point of view by the different characteristics of the silver pastes used in the process of the RLC filters. Indeed, the silver paste used for filter 2 has a different solvent than the one used for filter 1. The silver nanoparticles used in the case of filter 2 are diluted in an α-Terpineol solvent whereas in case of filter 1 the used solvent is Toluene. Taking into account that the boiling points of these different solvents are 217°C and 110°C respectively for α-Terpineol and Toluene and knowing that the final annealing is done for 30 mn at a temperature of 100°C, we can easily imagine that we will have a higher residual solvent proportion in case of filter 2. This has, as an immediate effect, the increasing of the serial resistance corresponding to the conductor losses. This will directly affect the cut-off frequency of the filter.

Figure 3 shows the representative model of processed RLC filter that has been used to simulate under Agilent ADS RF simulator [5] the S21 parameters.

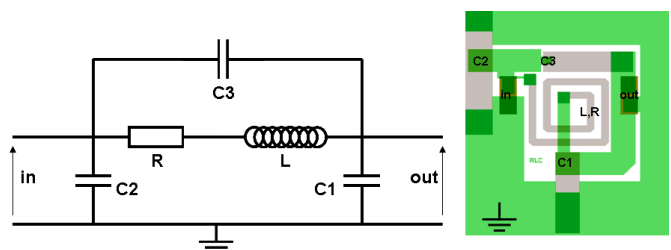

*Figure 3.* RLC simulated model and its correspondence with the circuit layout

We have estimated the capacitance values, assuming perfect electrodes, from the layout components using the formula:

$$C = \varepsilon_0 \cdot \varepsilon_r \cdot \text{area}/t \qquad (1)$$

where $\varepsilon_0$ is the vacuum permittivity, $\varepsilon_r$ the relative dielectric permittivity and t its thickness.

This simple calculation gives C1=C2=15 pF and C3 = 1.5 pF. The adjustment of the values of L and R has been performed by comparing the parameter S21 gain and phase between measurements and simulations. The values found for the inductance are Ls = 12 nH and Rs = 70 Ω for filter 1. Concerning filter 2 we have determined the same inductance value with a higher serial resistance Rs = 110 Ω.

We show on figure 4 the comparison between measurements and simulations using our simple RLC model.

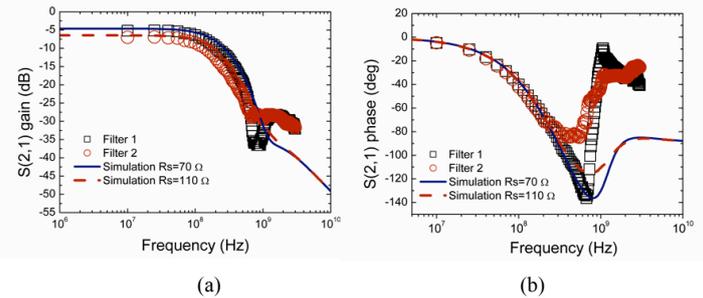

*Figure 4. Measured and simulated* transmission S21 parameters: gain (a) and phase (b) for both filters

We can see a good agreement between both gain and phase measurements and simulations. This confirms that the lower gain observed on filter 2 compared to filter 1 is mainly due to the additive parasitic serial resistance of the Ag nanoparticles paste.

## 3. Temperature sensors

Many studies on temperature sensors have been reported in literature in the field of silicon electronics applications. Nevertheless, in the organic semiconductors world few studies have been published [6] [7]. Our sensor is fabricated by the screen printing of resistive Positive Temperature Coefficient (PTC) and Negative Temperature Coefficient (NTC) inks on PEN and /or PET substrates. The final aim is to combine this fabricated temperature sensor with an RFID tag on the same flexible substrate. Figure 5 shows the variation of the sheet resistance $R_{sq}$ versus temperature of both PTC and NTC inks.

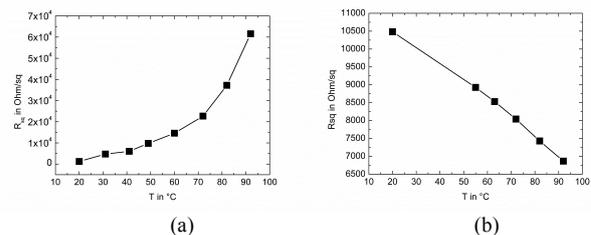

*Figure 5.* Sheet resistance variation versus temperature for PTC (a) and NTC (b) resistive inks

The extracted temperature sensitivities of both inks are about of 0.02-0.05 °C$^{-1}$ and 0.004-0.006 °C$^{-1}$ respectively for PTC and NTC resistive inks in the temperature range of 20-90 °C.

Using these resistive inks we have fabricated a set of Wheatstone bridges that are aimed to be used as temperature sensors by the screen-printing technique on PEN substrates.

Figure 6 shows respectively the foil (a) where several individual bridges were processed and a matrix array (b) dedicated to a temperature sensing application.

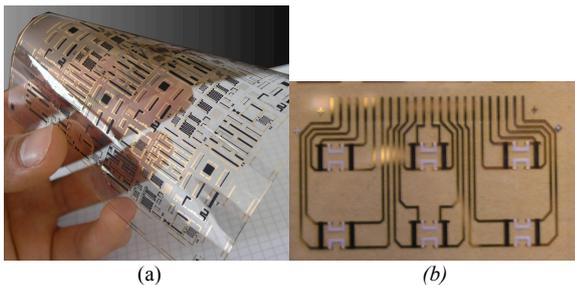

*Figure 6.* Image of the fabricated flexible foil (a) and the temperature sensor array (b)

The output voltages ($V_{out}$) of different temperature sensors have been measured between 20°C and 80°C with an input bias $V_{in}$= 4.8V. We present on figure 7 (a) and (b), respectively the output voltage ($V_{out}$) and the sensitivity of each temperature sensor versus temperature of the above fabricated array.

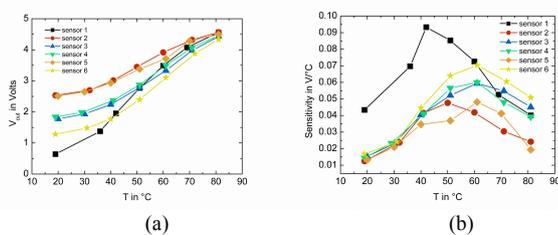

*Figure 7.* Measured output voltage (a) and sensitivity (b) of different temperature sensors)

We can see that we have a good agreement between the 6 different measured sensors. The sensitivity (S = $dV_{out}/dT$) of these temperature sensors was calculated by differentiating measured output voltage at the different applied temperatures. We obtain good sensitivity values for the different sensors. Disregarding the small mismatch between the sensors, the maximum average reached sensitivity of the different sensors is around 0.05 V/°C between 50°C and 70°C.

## 4. Conclusion

In this paper, we have described our manufacturing process of an organic RLC filter on PEN flexible substrate for RF applications and discussed the fabrication of temperature sensors in order to integrate them with an RFID tag on the same substrate. The use of PEN substrates and the printing techniques offer many advantages such as reducing the cost of the process flow without altering the electrical results. We observed that we are limited by the silver paste conductor characteristics in the case of RLC filters. The next step would be to work on the optimization of these silver pastes and the design of new RLC filter to increase their performance. Finally we showed in this paper a high sensitive temperature sensor fabricated by the same screen printing technique on flexible PEN substrate dedicated to an integration into an RFID system on the same flexible substrate.